\documentclass[%
 rsi, superscriptaddress,
 amsmath,amssymb,
 reprint,%
]{revtex4-2}

\usepackage{graphicx}
\usepackage{dcolumn}
\usepackage{bm}

\usepackage[utf8]{inputenc}
\usepackage[T1]{fontenc}
\usepackage{etoolbox}
\usepackage{caption}
\usepackage{subcaption}
\usepackage[export]{adjustbox}
\usepackage{tikz}
\usetikzlibrary{math}
\usetikzlibrary{calc}

\usepackage{siunitx}

\setcitestyle{super}

\newcommand\Pran{\mbox{\textit{Pr}}} 
\newcommand\Ra{\mbox{\textit{Ra}}} 
\newcommand\Rey{\mbox{\textit{Re}}} 
\newcommand\Nu{\mbox{\textit{Nu}}} 

\graphicspath{{./fig/}}

\makeatletter
\def\@email#1#2{%
 \endgroup
 \patchcmd{\titleblock@produce}
  {\frontmatter@RRAPformat}
  {\frontmatter@RRAPformat{\produce@RRAP{*#1\href{mailto:#2}{#2}}}\frontmatter@RRAPformat}
  {}{}
}%
\makeatother
\begin{document}

\preprint{AIP/123-QED}

\title[A Multipurpose Thermal Convection Setup to Study Turbulent Super Structures]{ A Multipurpose Thermal Convection Setup to Study Turbulent Super Structures}
\author{Hiufai Yik}
\affiliation{Max Planck Institute for Dynamics and Self-Organization, Am Fassberg 17, D-37077 G\"ottingen, Germany}
\author{Constantin Schettler}%
\affiliation{Max Planck Institute for Dynamics and Self-Organization, Am Fassberg 17, D-37077 G\"ottingen, Germany}
\author{Eberhard Bodenschatz}
\affiliation{Max Planck Institute for Dynamics and Self-Organization, Am Fassberg 17, D-37077 G\"ottingen, Germany}
\affiliation{Institute for the Dynamics of Complex Systems, University of Göttingen, Göttingen 37077, Germany}
\affiliation{Laboratory of Atomic and Solid State Physics, Cornell University, Clark Hall of Science, Ithaca, 14853, USA}
\author{Umesh Madanan}
\affiliation{Max Planck Institute for Dynamics and Self-Organization, Am Fassberg 17, D-37077 G\"ottingen, Germany}
\affiliation{Department of Mechanical Engineering, Indian Institute of Technology, Kanpur, 208016, India}
\author{Stephan Weiss}
\affiliation{Max Planck Institute for Dynamics and Self-Organization, Am Fassberg 17, D-37077 G\"ottingen, Germany}
\affiliation{Institute of Aerodynamics and Flow Technology,  German Aerospace Center (DLR),
Bunsenstrasse 10, D-37073 G\"ottingen, Germany}

\date{\today}

\begin{abstract}
A thermal convection apparatus has been designed to study turbulent super structures at high Rayleigh numbers and Prandtl numbers of the order of unity. This apparatus consists of a rectangular cell with a length of $3.50\,\mathrm{m}$, width of $0.35\,\mathrm{m}$, and variable height, which is fixed at $0.70\,\mathrm{m}$ for the present study. This cell is installed inside a $5.6\,\mathrm{m}$ long pressure vessel facility, known as \emph{G\"ottingen Uboot}, which can be filled with compressed gasses (air, helium, nitrogen, or sulfur hexafluoride) at pressures up to $19\,\mathrm{bar}$, enabling Rayleigh numbers up to $ \Ra \le 5\times 10^{12}$ and Prandtl numbers of approximately $0.7 \le \Pran \le 0.9$. The convection cell is bounded vertically by top and bottom plates consisting of a three-layer composite structure in which a thin Lexan plate is sandwiched between highly conductive aluminum plates. This allows for spatially resolved heat flux measurements. Each plate is subdivided into four longitudinal segments that can be independently temperature-controlled to enable homogeneous temperatures and the imposition of horizontal temperature gradients at both the top and bottom boundaries. While the bottom plate is electrically heated, the top plate's temperature is regulated using temperature-controlled circulating pressurized water. The apparatus is well suited for precise heat flux measurements, with the results obtained being in good agreement with those previously reported in the literature.
\end{abstract}

\maketitle

\section{Introduction}

Thermally driven convection flows in nature, such as those on planets and stars  are turbulent. \cite{schubert_turcotte_olson_2001} Consequently, fluid motion in such systems is very vigorous. These flows exhibit a wide range of length and time scales that far exceed the local scales of turbulence and are structured at large length and time scales. These structures are known as turbulent superstructures. \cite{ Stevens.ea2018,pandey2018turbulent} The fast range of scales renders their study using direct numerical simulations (DNS) in reasonable resolution impossible so far. Therefore, previous studies have investigated different aspects of convection through simplified and canonical settings. These studies aimed to determine power-law relations between dimensionless response variables and control parameters. The researchers presumed that these relations would extend to larger, real-world systems. \cite{ahlers_grossmann_lohse_2009}

One such heavily-researched canonical system is Rayleigh-B\'enard convection (RBC), where a
horizontal fluid layer of height, $H$, is heated from below and cooled from above.
\cite{Bodenschatz.ea2000,Kadanoff2001,Ahlers.ea2009C} For sufficiently small temperature difference
between the bottom and the top layer ($\Delta = T_b - T_t$), relevant thermophysical properties can
be assumed uniform and constant, with the exception of density ($\rho$) that depends linearly on
temperature. \cite{Oberbeck1879,Boussinesq1903} This is called the \emph{Oberbeck-Boussinesq} (OB)
approximation, with all thermophysical properties usually established at the mean temperature
$T_m=(T_b + T_t)/2$. Here, the thermal driving is  expressed in its dimensionless form by the Rayleigh number:
\begin{equation}
\Ra = \frac{\alpha g\Delta H^3}{\nu\kappa}\mbox{,}
\end{equation}
where $\alpha$ denotes the isobaric volumetric expansion coefficient, $g$ is the gravitational acceleration, $\nu$ is the kinematic viscosity and $\kappa$ is the thermal diffusivity. The validity of the Oberbeck–Boussinesq (OB) approximation for compressed gases has been investigated in detail. \cite{shishkina_weiss_bodenschatz_2016,weiss_he_ahlers_bodenschatz_shishkina_2018}

The second intrinsic control parameter in RBC, which quantifies the relative significance of viscous and thermal diffusion is the Prandtl number, given by:
\begin{equation}
\Pran=\nu/\kappa\mbox{.}
\end{equation} 

In addition to these two control parameters, the geometry of the convection cell plays a significant
role. This is usually captured with the help of aspect ratio ($\Gamma = D/H$), defined as the ratio
of the lateral dimension, $D$, and cell height, $H$. The aspect ratio controls the influence of the sidewalls on the flow. 

The state of RBC is usually characterized by two global response parameters, namely the Nusselt
number 
\begin{equation}\label{eq:Nu}
\Nu=\frac{qH}{\lambda \Delta} \mbox{,}
\end{equation} 
and the Reynolds number 
\begin{equation}\label{eq:Re}
\Rey = \frac{UH}{\nu}\mbox{.}
\end{equation}
Here, $\lambda$ refers to the thermal conductivity of the fluid and $U=\sqrt{\langle u^2 + v^2 + w^2\rangle}$ being the average velocity of the flow. 
Nusselt number compares the volume and time averaged heat flux $q$ with the heat flux in the same system in absence of any convection (or, bulk fluid motion). While the global $\Nu$ can be measured through experiments, computing local properties requires the entire velocity and temperature field to be resolved, which is quite demanding. \cite{barta_volk_bauer_wagner_mommert_2025}

With an increase in \Ra, the range of relevant length scales in the flow also increases.
Subsequently, the ratio of the height scale ($H$) to the smallest (the Kolmogorov scale,
$\eta$) follows a power law relation of $H/\eta \sim \Ra^{1/3}$. For instance, in the Earth's
atmosphere \cite{Chilla.ea2012}, $\Ra$ can reach $\sim 10^{18}$,  creating turbulence with length
scales ranging from a few centimeters to many kilometers. Such systems cannot yet be comprehensively
studied either experimentally or simulated numerically. The current state-of-the-art 3-dimensional
simulations can simulate an RBC system with $\Ra$ values up to $10^{12}$ in a cylinder of aspect
ratio $\Gamma=1$ while fully resolving for all relevant parameters. \cite{stevens2020toward} However, for a large-scale, highly-turbulent system that is comparable to some of the aforementioned natural systems, the required computational power will be many orders of magnitude higher than what is currently feasible.

Since natural conditions cannot be emulated in the laboratory setting or through simulations, the
alternative is to establish scaling relations for the dependencies of \Rey(\Ra,\Pran)\ and
\Nu(\Ra,\Pran) or other simplified models to help extrapolate laboratory results to real-world
scenarios. \cite{Malkus1954, kraichnan1962turbulent, grossmann2000scaling, grossmann2004fluctuations,
grossmann2011multiple} The primary prerequisite to establish such scaling relations is the capacity
to measure system responses over sufficiently large ranges of \Ra\ and \Pran. However, one of the
major challenges associated with thermal convection experiments is achieving large $\Ra$ with
sufficiently small $\Delta$ to adhere to the OB approximation within the domain. Despite \Ra\ having
a steep scaling with height ($\Ra \propto H^3$), even the largest convection cells built till date
(the ``Barrel of Ilmenau’’, \cite{du2007structure} a cylindrical cell filled with air having a
diameter of 7.15\,m and a height of up to 6.3\,m) is capable of realizing $\Ra$ values only up to
$10^{12}$, which is several orders of magnitude lower than \Ra\ in natural systems. Therefore,
different approaches have been adopted to achieve $\Ra$ values beyond $10^{12}$. These approaches
include using a fluids with large thermal expansion coefficient ($\alpha$) and extremely low
viscosity such as cryogenic helium \cite{urban2010helium, niemela2000turbulent,
chavanne1997observation, castaing1989scaling} or gases with large pressures (or, densities) and hence small $\nu$
and $\kappa$ \cite{madanan2020high}. Using gases as a working fluid also has the advantage that one can control their thermophysical
properties over a large range by varying the operating pressure. A popular choice for such a gas is
sulfur hexafluoride (SF$_6$), \cite{ashkenazi1999high,ahlers2012heat} with almost six folds higher
density than air. Although cryogenic helium \cite{niemela2000turbulent} has also been used similarly to attain Rayleigh
numbers up to $Ra=10^{17}$, the reliability of the obtained results was 
challenged due to deviations from the OB approximation (NOB), \cite{wu1991non, ahlers2007non,
urban2014heat, weiss2018bulk, yik2020turbulent}, especially since heat transport under NOB conditions
can be significantly greater \cite{yik2020turbulent} than for the corresponding OB case. Pressurized
SF$_6$ has been successfully employed in recent years for highly turbulent RBC experiments in
cylindrical cells with diameters of $D=1.1$\,m and aspect ratios $\Gamma = 1/2$ and $\Gamma = 1$.
There Rayleigh numbers of up to $\Ra = 10^{15}$ were achieved, and a regime of stronger heat transport (the ultimate regime) was observed. \cite{He.ea2012A}

Typically, experimental and computational studies which aim to investigate global heat fluxes in turbulent RBC are performed in cells with
$\Gamma$ close to unity, as it is assumed  that \Nu\ is nearly independent of $\Gamma$
for $\Gamma \gtrapprox 1$. \cite{Stevens.ea2018,Shishkina.2021} Convection in such cells usually
takes the form of a single large-scale circulation (LSC) roll that spans the entire cell so that
warm fluid rises on one side and cold fluid sinks along the other.\cite{brown2007large} However, some recent studies on
turbulent RBC with larger aspect ratios
\cite{Bailon-Cuba.ea2010,Stevens.ea2018,pandey2018turbulent,krug2020coherence,weiss.ea2023} have
revealed that multiple coherent rolls can exist even under highly turbulent conditions. These are known as turbulent superstructures.

An improved understanding of the self-organization of large-scale convection patterns is required to obtain deeper insight into the overall flow dynamics in large-aspect-ratio systems. Such understanding is particularly important, as turbulent superstructures represent the prevailing state in natural convective flows. A review of the relevant literature indicates that investigations addressing the dynamics of the large-scale circulation (LSC) in large-aspect-ratio systems at high $\Ra$ remain limited. To overcome these limitations, the Long High Pressure Convection Facility (LHPCF) has been designed and constructed as a rectangular convection cell with an adjustable height, thereby enabling experiments over a broad range of high Rayleigh numbers. Using this facility, experiments on Rayleigh–Bénard convection (RBC) can be performed for $\Ra$ values up to $5\times10^{12}$ at an aspect ratio of $\Gamma_x=5$, and up to $10^9$ at $\Gamma_x=35$. For horizontal convection, in which the driving temperature or density difference is imposed along a horizontal boundary, Rayleigh numbers as high as $10^{16}$ can be achieved. The facility is further equipped to allow spatially resolved heat-flux measurements at both the top and bottom plates. In addition, optical access is provided through transparent plexiglass sidewalls, enabling flow diagnostics such as particle image velocimetry (PIV) and high-density Lagrangian particle tracking to be conducted.

\section{Construction}

\subsection{Overview}

\begin{figure}[h]
	\centering
    \includegraphics[width=0.49\textwidth]{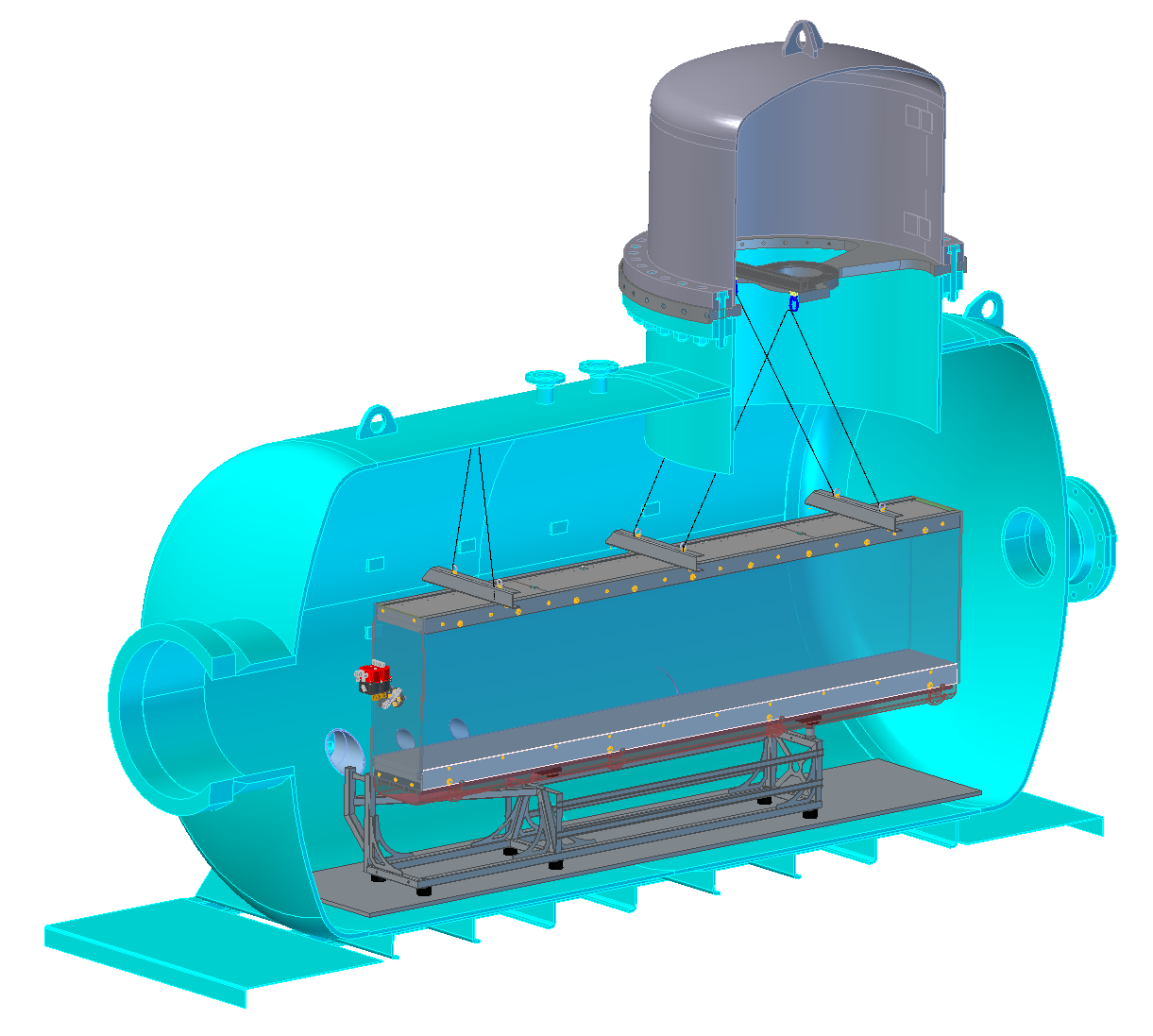}
	\caption{Overview of LHPCF consisting of a rectangular RBC cell placed inside the G\"ottingen Uboot. 
            }
	\label{fig:overview}
\end{figure}

\begin{figure*}[htbp]
	\centering
	\includegraphics[width=\textwidth]{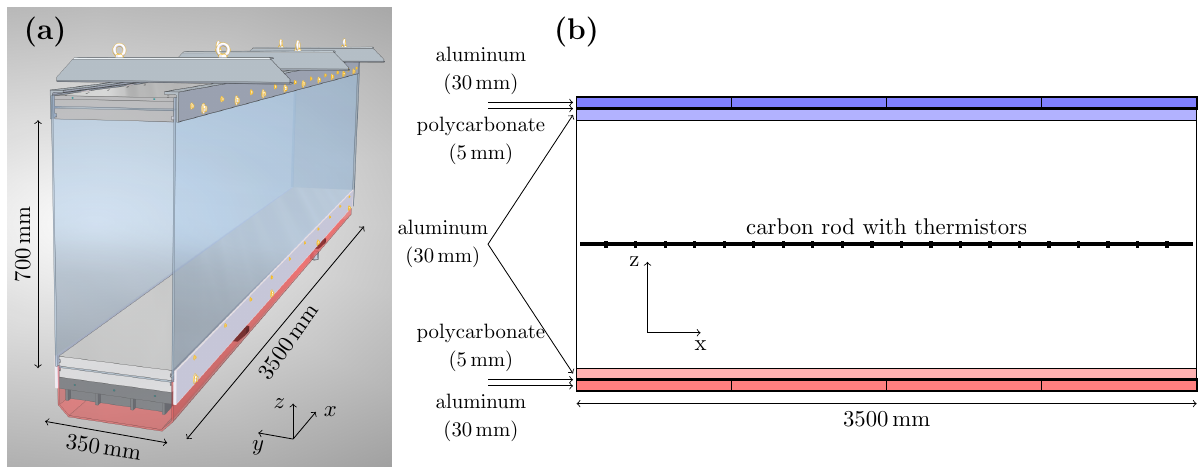}
	\caption{(a) CAD rendering of the entire RBC cell and (b) Side-view sketch of the RBC cell. Note: side view is scaled-down in the $x$-direction by a factor of two for better readability.}
	\label{fig:setup}
\end{figure*}

Figure ~\ref{fig:overview} illustrates a rendered image of the LHPCF. A convection cell is
constructed and installed inside the high-pressure vessel known as the \emph{``Uboot of
G\"ottingen’’}. This facility is located at the Max Planck Institute for Dynamics and
Self-Organization, G\"ottingen, and has been previously used for convection experiments at large
\Ra. \cite{Ahlers.ea2009B,He.ea2012,wedi.ea2022} A detailed description is
available in a publication by Ahlers et al. \cite{Ahlers.ea2009B} The Uboot consists of a 5,6\,m
long horizontal steel cylinder with a diameter of $2.5\,$m  and a 1.5\,m high turret. Although the
Uboot can be filled with a variety of non-corrosive gases, such as helium or nitrogen, it is
typically operated either with air at atmospheric pressure (\Pran $= 0.7$) or sulfur hexafluoride
(SF$_6$) at different pressures of up to $19\,$bar (\Pran $= 0.9$), the maximum operating pressure
of this facility. For operation with SF$_6$, a dedicated gas-handling infrastructure is available,
including $20\,$t of SF$_6$ stored in reservoirs, as well as vacuum pumps, compressors, and a
distribution system. This infrastructure facilitates efficient evacuation, filling, and
pressurization of the Uboot.

To maintain a constant and spatially uniform temperature within the Uboot, an industrial heater with
a power rating of 9000\,W is installed. Heated water is driven through tubes wrapped around the
outer circumference of the cylinder with the aid of a circulation pump. To minimize thermal
interactions within the testing environment, the Uboot is insulated with a $\approx 50\,$mm-thick
layer of foam. The heater temperature is regulated using a computer-controlled feedback system,
which ensures that the Uboot temperature ($T_u$) is maintained at up to $15\,$K above the laboratory
temperature. However, $T_u$ is typically set to no more than $5\,$K above the laboratory
temperature, which is sufficient to compensate for temperature fluctuations in the surrounding
environment. The heating power is adjusted dynamically based on temperatures measured within the Uboot. In addition, gas inside the Uboot is continuously mixed by fans to ensure a homogeneous temperature distribution throughout the vessel, excluding the interior of the convection cell.

Figure ~\ref{fig:setup} schematically illustrates the convection cell. This cell has a rectangular
horizontal cross-section with length, $L=3.5$\,m, width, $W=0.35$\,m, and height, $H=0.7$\,m,
providing aspect ratios of $\Gamma_x=L/H=5$ and $\Gamma_y=W/H=0.5$. The sidewalls and front and rear
walls of the cell are fabricated using Plexiglas sheets with thicknesses of $5\,$mm (sidewalls) and
$10\,$mm (front and rear walls), respectively. The convection cell also consists of a bottom
construction resting on the floor of the Uboot and a top assembly suspended from the ceiling of the
Uboot. These side, front, and rear Plexiglas sheets are mounted between the top plate assembly
(described in Sec.~\ref{sec:top-plate}) and the bottom plate to prevent unwanted gas exchange
between the cell's interior and the surrounding environment. An electrically actuated valve is
installed at the center of the front wall at mid-height (red in Fig.~\ref{fig:overview}). This valve
is opened to establish pressure equilibrium between cell interior and the Uboot ambience during
filling, evacuation, or temperature adjustment of the system. This valve is kept closed during
measurement to prevent any gas exchange between the cell and the ambience. 

\subsection{The bottom plate construction}

The bottom plate assembly consists of the bottom plate sandwich, a thermal shield, and a supporting kinematic table. 
The  assembly (see the red colored part in Fig.~\ref{fig:setup} (a) and (b)) is constructed as a three-layer sandwich, consisting of a $l_{pc}=5\,$mm thick Lexan (polycarbonate) sheet with a thermal
conductivity, $\lambda_\textrm{pc} = 0.2,\mathrm{W/(m\cdot K)}$, bonded between two $30\,$mm thick
aluminum plates (alloy EN AW 5083 manufactured by Gleich, with $\lambda_\textrm{Al} =
110$–$130,\mathrm{W/(m\cdot K)}$). These plates are glued together using degassed Stycast 1266 epoxy. The two aluminum layers, henceforth referred to as the bottom-plate bottom (\emph{"bb"}) and bottom-plate top (\emph{"bt"}), are shown in Fig.~\ref{fig:setup}(b) in dark red and light red colors, respectively. While the $bt$ consists of a single $3.5\,$m long aluminum plate, the $bb$ is composed of four $0.875\,$m long, independently heated segments. This is designed to generate lateral temperature gradients to drive large-scale flow patterns. 
Furthermore, in this way the risk of air entrapment within the adhesive layer when glueing the
plates together is strongly reduced.

\begin{figure}[htbp]
	\centering
	\includegraphics[width=\columnwidth]{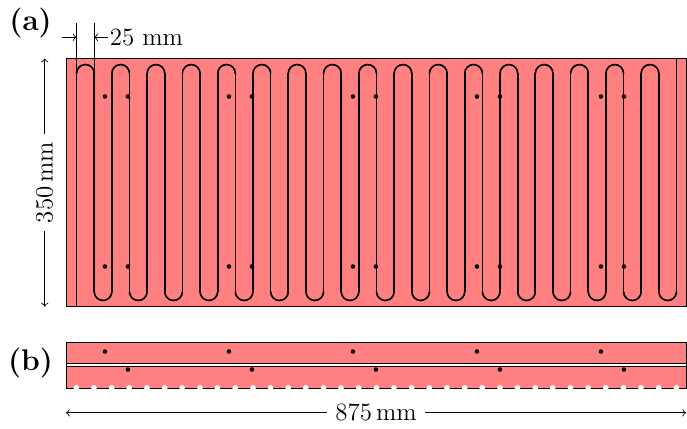}
	\caption{(a) Bottom-view of a single segment of the bottom plate with grooves and heating cable and (b) Side-view of the same segment in the sandwich. Note: black dots mark the thermistor locations.}
	\label{fig:bottom-plate}
\end{figure}

\begin{figure}[htbp]
	\centering
	\includegraphics[width=\columnwidth]{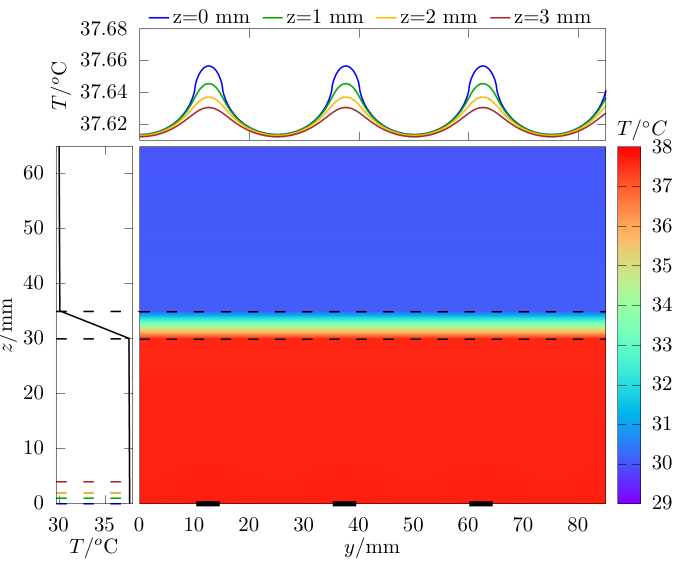}
	\caption{Simulated temperature field of a vertical slice of the bottom plate under constant heat
	flux from its top surface, which is provided by localized heating wires at the bottom (black short bars). Note: (i) horizontal black dashed lines denote the interfaces between the aluminum and the Lexan plates, (ii) black solid line on the left sub-plot shows the average temperature profile as a function of height, and (iii) dashed colored lines in the left sub-plot indicate the positions at which horizontal temperature profiles are shown in the top sub-plot (refer to legend for exact positions). 
	}
	\label{fig:temp-field-bottom}
\end{figure}

Figs ~\ref{fig:bottom-plate} (a) and (b) show a single $0.875\,$m long segment of the bottom plate
assembly. The $bb$ is heated electrically using a $47.54\,$m long resistive heating wire with a
total resistance of $6.6\,\Omega$, which is embedded within $5\,$mm deep grooves milled into the
lower surface of the bottom plates and fixed with glue (Stycast epoxy). In this way, good thermal contact
between the heating wire and the plate is ensured. The grooves for the heating wire are milled in a
serpentine pattern, with adjacent channels spaced $25\,$mm apart (see Fig. ~\ref{fig:bottom-plate}
(a)). This spacing is sufficiently small to maintain a homogeneous temperature distribution at $bt$. In order to verify the homogeneity, the
heat diffusion equation is solved numerically and the resulting temperature field is shown in Fig.~\ref{fig:temp-field-bottom}. A constant heat flux boundary condition is applied at the top surface ($q=\SI{300}{\watt\per\square\metre}$), corresponding to the heating provided by \SI{5}{\milli\metre} wide heating elements at the bottom surface (depicted using short black lines in the figure). Adiabatic boundary conditions are assumed outside the heating elements, and only the average temperature at the $bt$ is fixed at \SI{30}{\degreeCelsius}. The resulting temperature field showed no significant inhomogeneities. Horizontal temperature variations induced by the heating elements decay within a few millimeters above the lower boundary (top sub-plots in Fig.~\ref{fig:temp-field-bottom}), whereas the vertical temperature profile (left sub-plot in Fig.~\ref{fig:temp-field-bottom}) indicates that temperature drop occurs primarily across the Lexan layer with negligible temperature gradient within the aluminum plates.

The heating wires for each of the four segments are powered by power supply units (Agilent 6675A for
the two outer segments and Keysight E3634A for the two inner segments) controlled via a computer
interface. $bb$ and $bt$ temperatures are measured using $78$ thermistors positioned along two
columns, marked by black dots in Fig.~\ref{fig:bottom-plate}. 

\begin{figure}[htbp]
	\centering
	\includegraphics[width=\columnwidth]{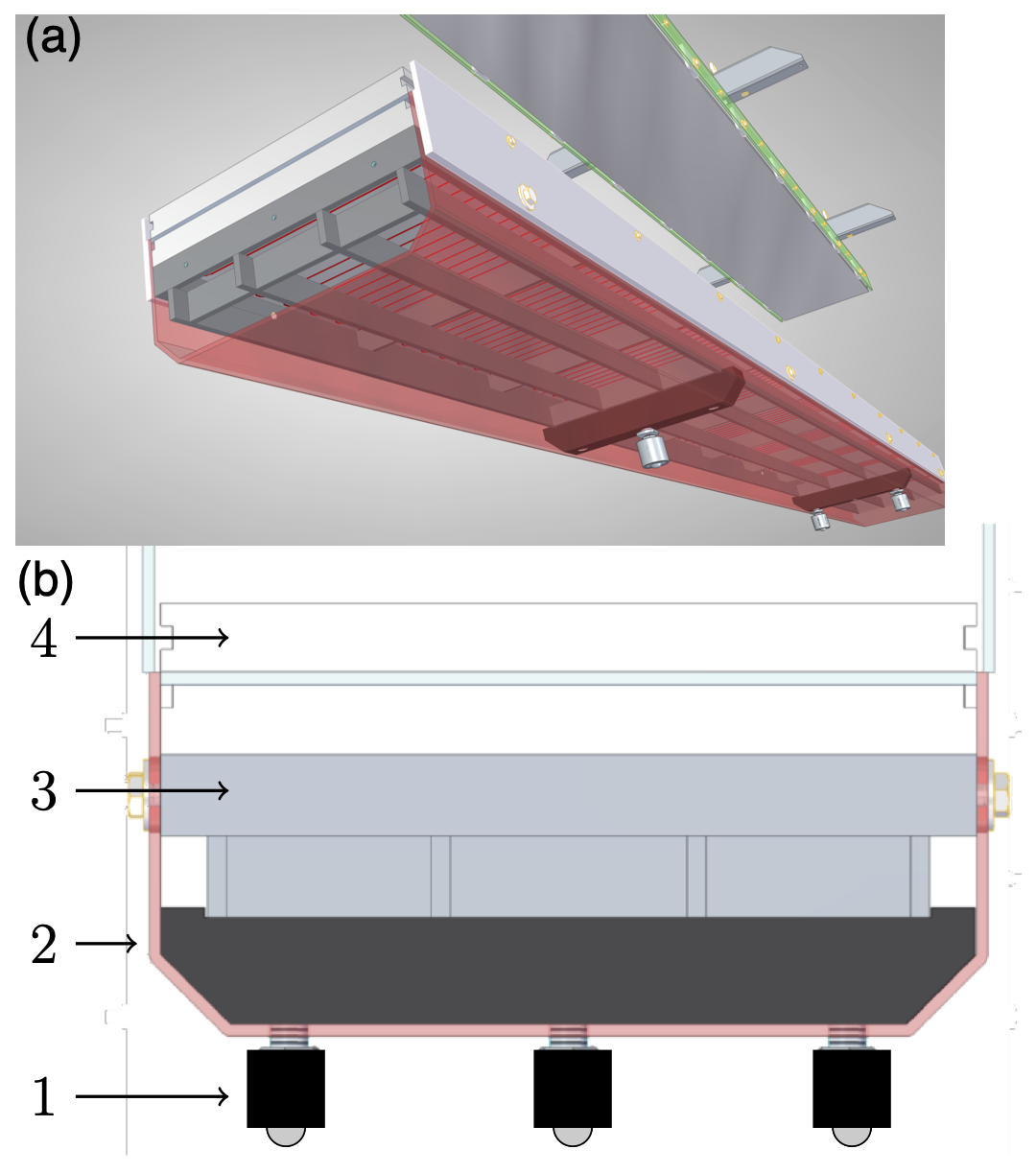}
	\caption{(a) Overview of the supporting grid with thermal shield (colored red). (b) Cross-section of the entire bottom plate construction. Note: the numbers denoted in (b) mark: (1) supporting ball transfer units as feet to rest on the kinematic table, (2) thermal shield, (3) supporting grid, and (4) bottom plate sandwich.}
	\label{fig:bottom-plate-overview}
\end{figure}

The bottom plate sandwich rests on a supporting stainless-steel grid, as shown in
Fig.~\ref{fig:bottom-plate-overview}. This grid consists of a staggered arrangement of $8\,$mm thick
and $35\,$mm high stainless-steel sheets, configured to minimize thermal contact points and reduce
conductive heat losses in the downward direction. Similarly, all voids within this grid are filled
with foam and thin rubber layers are inserted at all contact points between the bottom plate and the
supporting grid. A thermal shield (colored red in Fig.~\ref{fig:bottom-plate-overview} (a) and (b))
is also installed beneath the supporting grid in the form of a bent metal sheet to further reduce
heat losses. This shield is additionally insulated from the Uboot environment using a foam layer,
and its temperature is maintained equal to that of the $bb$ via electrical heating wires attached to
it.

The entire bottom plate assembly is mounted on a kinematic table inside the facility and aligned to within 0.02\,$^\circ$ relative to gravity. The weight of the supporting grid is supported by three ball transfer units resting on the kinematic table (see Fig.~\ref{fig:bottom-plate-overview}b).

\subsection{The top plate assembly}\label{sec:top-plate}

\begin{figure}[htbp]
	\centering
	\includegraphics[width=\columnwidth]{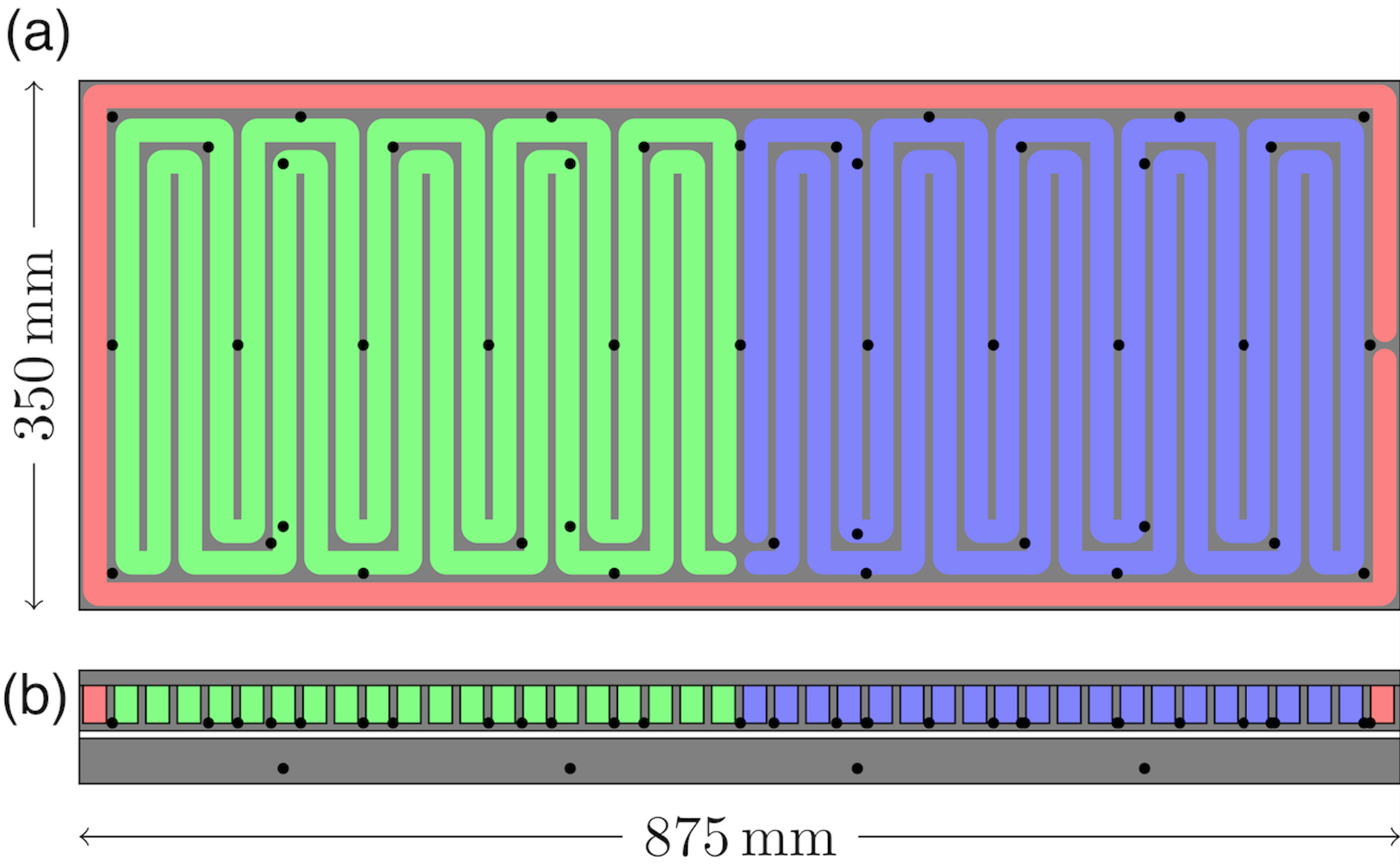}
	\caption{(a) Top-view of the top plate with grooves for the cooling water. Note: the grooves marked in green, blue, and red mark independent cooling circuits. (b) Cross-section of the top plate showing channels for cooling water, aluminum plates (colored gray), and the Lexan plate sandwiched between the two (colored white). Note: black dots mark the thermistor locations.
	}
	\label{fig:top-plate}
\end{figure}

The top plate assembly is also constructed as a three-layer sandwich, with a $l_{pc}=5\,$mm thick Lexan sheet sandwiched between two $30\,$mm thick aluminum plates using degassed Stycast 1266 epoxy. The bottom layer of this assembly is the fluid-facing aluminum plate (top-plate bottom or \emph{tb}) manufactured as a single $3.5\,$m long plate, whereas the upper layer (top-plate top or \emph{tt}) is composed of four $0.875\,$m long independently heated smaller segments. Fig.~\ref{fig:top-plate} (a) and (b) respectively illustrate the top- and cross-sectional views of the top plate assembly. 

The $tt$ is cooled using temperature-controlled water circulating through grooves measuring $15\,$mm in width and $17\,$mm in depth (shown in red, green, and blue in Fig.~\ref{fig:top-plate} b). These grooves are milled into the upper surface of the $tt$, covered with an additional \SI{10}{\milli\metre} thick aluminum plate, and then joined through diffusion welding, i.e., they are heated to a temperature close to the solidus temperature of aluminum and pressed together in a vacuum oven. These grooves form three independent water circuits within each plate segment (marked in green, blue, and red in Fig. ~\ref{fig:top-plate}). These three water circuits are required to minimize temperature gradients within the plate since the cooling water warms while absorbing heat during its passage through the channels. Two of these circuits serve the left and right halves of each segment, whereas the third circuit (located near the edges; marked red) compensates for additional heat losses through the sidewalls. Cooling water is circulated through the circuits at a flow rate of $0.5\,$l/s in a closed-loop system and its temperature is regulated by a heat exchanger connected to multiple chillers (ThermoScientific AC-150; one for each loop). These chillers provide a combined cooling power exceeding \SI{2}{\kilo\watt}; this capacity can be readily increased through the addition of further chillers or by employing units with higher cooling power.

\begin{figure}[htbp]
	\centering
	\includegraphics[width=\columnwidth]{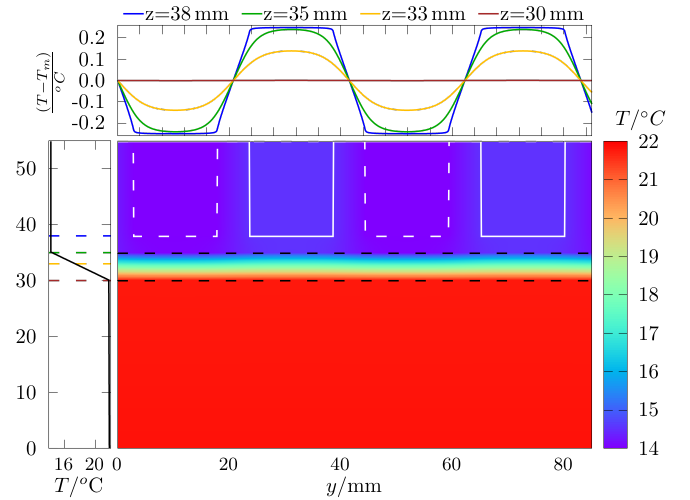}
	\caption{Simulated temperature field of a vertical slice of the top plate. Note: (i) this
	assumes constant heat flux at the bottom, fixed temperatures in the water leading channels, a
	temperature difference of \SI{0.5}{K} for cooling water entering/leaving the neighboring
	channels. (ii) horizontal black dashed lines denote the interfaces between the aluminum and the
	Lexan plates, (iii) white solid and dashed lines indicate channels for the cooling water
	containing cool water and heated water, respectively, (iv) the black solid line on the left sub-plot shows the average temperature profile as a function of height, (v) dashed colored lines in the left sub-plot indicate the positions at which horizontal temperature profiles are shown in the top sub-plot (refer to legend for exact positions). 
	}
	\label{fig:temp-field-top}
\end{figure}

Fig.~\ref{fig:temp-field-top} shows a simulated temperature field of the top plate assembly. For
this simulation, it is assumed that the temperature of the channel carrying the incoming cooling
water is approximately \SI{0.5}{K} lower than that of the neighboring return-flow channel.
Furthermore, a uniform heat flux of \SI{300}{\watt\per \square \metre} is assumed from the fluid to
the lower surface of the top plate. It should be noted here that the assumed heat flux of
\SI{300}{\watt\per \square \metre} is realistic, whereas the assumed temperature difference of
\SI{0.5}{K} represents a considerable overestimation. In practice, the temperature difference between the inlet and outlet of the cooling water does not exceed \SI{0.1}{K}, even at the highest \Ra. Owing to the large dimensions and depth of the cooling channels, significant temperature variations (approaching \SI{0.5}{\kelvin}) are still present at the interface between the $tt$ and the Lexan layer, although thermal diffusion noticeably smooths the temperature gradients. With increasing distance into the Lexan layer, the temperature field becomes progressively more homogeneous. Consequently, no discernible temperature inhomogeneities remain at the interface between the Lexan layer and the $tb$.

\begin{figure}[htbp]
	\centering
	\includegraphics[width=\columnwidth]{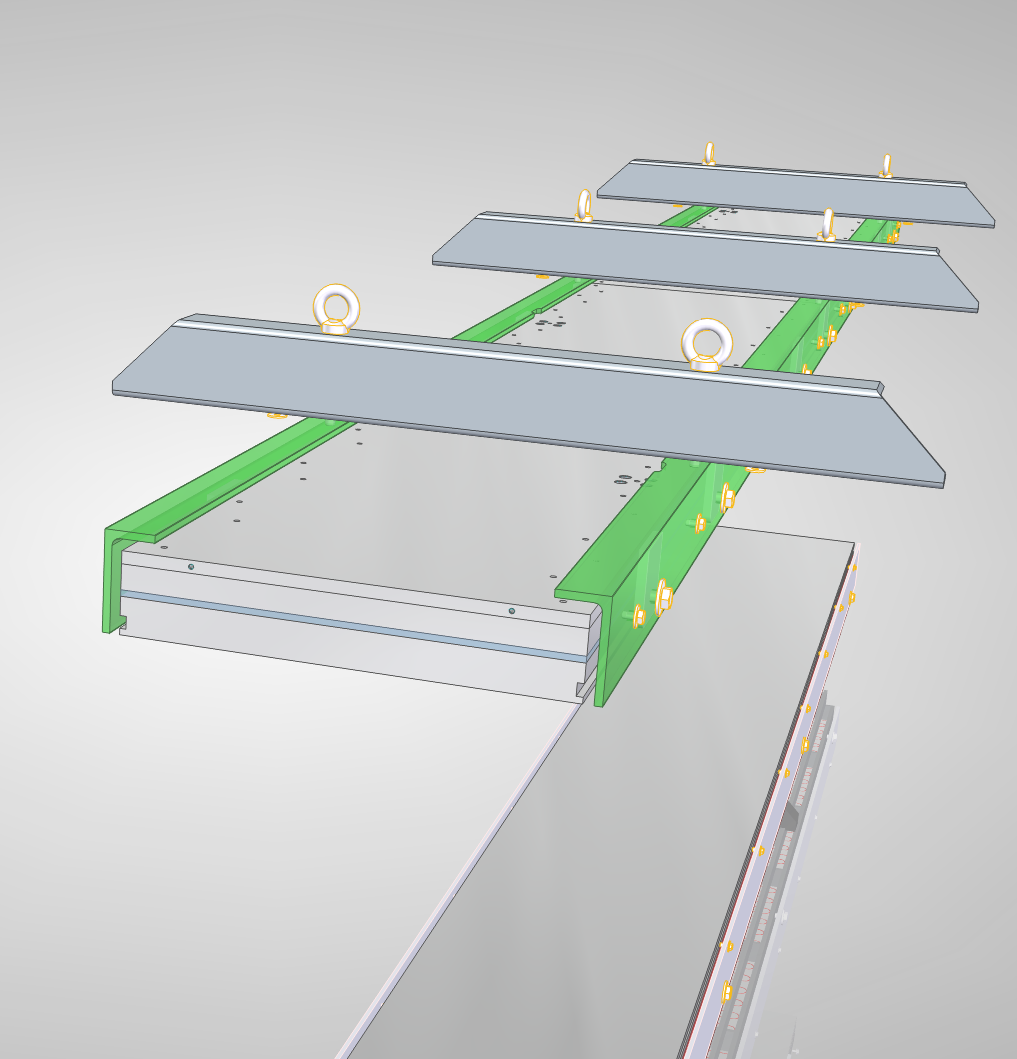}	
	\caption{Top plate construction showing the top plate sandwich, L-profile, and the horizontal support bars to support the top plate.}
	\label{fig:top-plate-overview}
\end{figure}

The top plate assembly is connected to an L-shaped profile (marked green in
Fig.~\ref{fig:top-plate-overview}) by means of 18 low-thermal-conductivity polyamide screws (M10).
An $8\,$mm gap is maintained between the L-profile and the top plate assembly using small plastic
spacer plates located at the screw positions. These design features are implemented to minimize heat
transfer from the warmer Uboot ambience to the top plate assembly and in this way ensuring
homogeneous temperature conditions. Furthermore, the sidewalls are also attached to the same L-shaped profile. This provides structural rigidity against gravitational bending and supports a total weight of approximately $240\,$kg while being suspended from the ceiling of the Uboot, as shown in Fig.~\ref{fig:top-plate-overview}. To this end, three horizontal bars are attached to the L-profile. Steel cables capable of supporting the entire weight of the top plate assembly are connected to these bars via eye bolts. These cables are guided over three pulleys mounted to the ceiling of the Uboot, as shown in Fig.~\ref{fig:overview}. The cable lengths can be finely adjusted using turnbuckles, allowing the top plate to be leveled to within 0.03\,$^{\circ}$ with respect to gravity. In this configuration, the top plate assembly is suspended at three points in space, facilitating straightforward adjustment of the cell height. By replacing the sidewalls, different cell heights and, consequently, different \Ra\ values can be realized with relatively little effort.

\subsection{Sensors}

Temperature distribution within the top and bottom plate assemblies is determined using 264 NTC
thermistors (B57861S0104, manufactured by EPCOS AG). These thermistors are calibrated in a dedicated
calibration box outside the Uboot against a commercial high-precision platinum temperature sensor
(PT-103-70H, Lake Shore Cryotronics) over a temperature range of 10\,$^\circ$C to 70\,$^\circ$C,
resulting in a relative error of $0.015\,$K. These thermistors are installed in the $tt$, $tb$, $bt$, and $bb$ aluminum sections, as indicated by the black dots in Fig.~\ref{fig:bottom-plate} and Fig.~\ref{fig:top-plate}. The thermistors located in the $bt$ and $tb$ are inserted into blind holes from the sides. In contrast, the thermistors in the $tt$ and $bb$ are inserted vertically from the top and bottom surfaces, respectively.

The thermistor resistances are measured using six Agilent Keysight 34970A multimeters, which are
triggered in parallel at intervals of $9\,$s. These multimeters are connected to a computer, where
temperatures are calculated from the measured resistance values using appropriate calibration
relationships. This computer also controls all power supplies and chillers based on values obtained
from the proportional-integral-derivative (PID) feedback control loops. Since temperatures of the
top and bottom plate assemblies are determined as functions of both time and the $x$-coordinate,
temperatures recorded by each thermistor are interpolated in time. Therefore,
temperature values for all thermistors are obtained at a common point in time, despite the
consecutive acquisition of the 60 channels of a single multimeter, which introduces time differences
of up to $10\,$s between the first and last thermistor measurements. As two thermistors are
typically located at a given $x$-position but at different y-positions (see
Fig.~\ref{fig:bottom-plate}), their measured temperatures are averaged to get the temperature profile ($T(x)$).

After some initial measurements, an array of 20 thermistors are installed along the $x$-direction at the mid-height and mid-width of the cell. These thermistors are mounted at an equal spacing along a $3.5\,$m long carbon-fiber rod with an outer diameter of $7\,$mm, as shown in Fig.~\ref{fig:setup}. Although both the thermistor response time and the measurement frequency are relatively slow (approximately $1.3\,$s in pressurized SF$_6$ at $18\,$bar), the array provides valuable information on the orientation and dynamics of the large-scale circulation structure.

\subsection{Heat flux measurements}

The sandwich structure of the top and bottom plate assemblies, together with the spatially resolved temperature measurements, enables the local heat flux at both the top and bottom plates to be determined as a function of the $x$-coordinate:
\begin{equation}
q_{t,b}(x)=\frac{\lambda_{pc}\Delta_{t,b}(x)}{l_{pc} }\mbox{.}
\end{equation}

Here, the indices $t$ and $b$ denote the top and bottom plate assemblies, respectively, and $\Delta_{t,b}(x)$ represents the temperature drop across the Lexan plate.

Furthermore, since
$\frac{\Delta_{Al}}{\Delta_{pc}}=l_\textrm{Al}\lambda_\textrm{pc}/\lambda_\textrm{Al} l_\textrm{pc}
\approx 0.01$, the temperature drop across the aluminum plates is considered negligible compared to that across the Lexan plates.

\section{Test Measurements}
\subsection{Spatial homogeneity}
It is evident from Figs.~\ref{fig:temp-field-bottom} and ~\ref{fig:temp-field-top} that temperature
distribution at the fluid-plate interface remains spatially homogeneous under conditions of constant
heat flux between the fluid and vertical boundaries. These results, however, only demonstrate the
negligible influence of cooling-channel structure at the top plate assembly and the heating-wire
arrangement at the bottom plate assembly. Under turbulent conditions, heat flux near the plate is
spatially inhomogeneous and varies in time. Two-dimensional simulations
\cite{Zhu.ea2018A,reiter.ea2021} indicate that heat fluxes strongly differ between the
plume-ejecting and the plume impacting regions. Enhanced heat flux in the flow in some regions causes surface temperature at the bottom (or, top) plate assemblies to decrease (or, increase). Consequently, a larger vertical temperature difference across the sandwich plate, $\Delta_{t,b}(x)$, is established, leading to increased values of the measured local heat flux ($q_{t,b}(x)$).

\begin{figure}[htbp]
	\centering
	\includegraphics[width=\columnwidth]{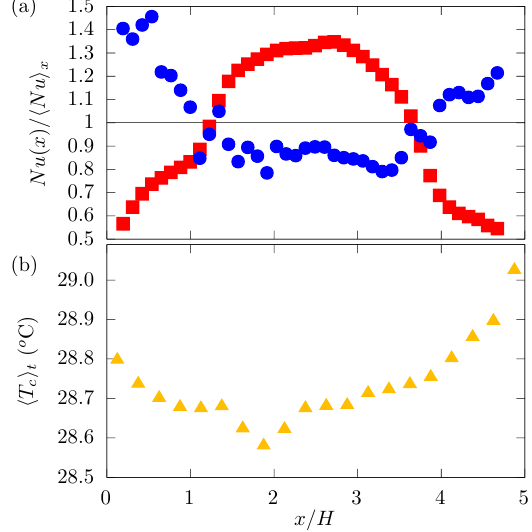}
	\caption{Spatially resolved heat flux for a state where warm fluid rises at the sides and cold
	fluid sinks in the middle. (a) Spatially resolved Nusselt number measured at the top (blue
	bullets) and bottom (red squares) plates as functions of $x$. (b) Temperature distribution at
	midheight. Note: measurements were taken at $\Delta=5.0$\,K and P=2\,bar ($\Ra=5.5\times
	10^{10}$), and averaged over $\approx 50$ free fall time units.}
	\label{fig:Nu-x-top-bot}
\end{figure}

$q_{t,b}$ values are found to vary substantially with the horizontal coordinate $x$. The corresponding local Nusselt numbers at the bottom (red solid squares) and top (blue solid circles) plate assemblies are estimated using $\Nu_{t,b}(x)=q_{t,b}(x)H/\lambda \Delta$ and normalized with their horizontal averages $\langle \Nu_{t,b}\rangle_x$, before plotting in Fig.~\ref{fig:Nu-x-top-bot}(a). At the bottom plate assembly, the local heat flux in the central region of the cell ($x\approx 2.5H$) exceeds the horizontal average by more than 30\%, whereas values about 40\% of the average are observed closer to the boundaries. An opposite trend is observed at the top plate assembly. This behavior is consistent with the large-scale circulation pattern, as regions where thermal plumes are ejected from the bottom plate correspond to regions where plumes impinge on the top plate.

These results are also compared with the horizontally resolved mean temperature at mid-height,
$\langle T_c\rangle (x)$. A lower temperature is observed in the central region of the cell, whereas
higher temperatures are measured near the lateral boundaries. This temperature distribution suggests
the presence of two counter-rotating circulation rolls, with warm fluid ascending along the
sidewalls and cold fluid descending in the cell center. Accordingly, plume-impacting regions are
located at the center of the bottom plate assembly and near the lateral sides of the top plate
assembly. Enhanced heat flux is observed in these regions. In contrast, plume-ejecting regions are
associated with a reduced heat flux and are located near the lateral sides of the bottom plate
assembly and in the central region of the top plate assembly. This observation is in agreement with
behavior reported for DNS of two-dimensional RBC systems for similar \Ra\ and \Pran\ \cite{reiter.ea2021}.

\subsection{Effect of ambient temperature}

\begin{figure}
    \centering
    \includegraphics[width=\linewidth]{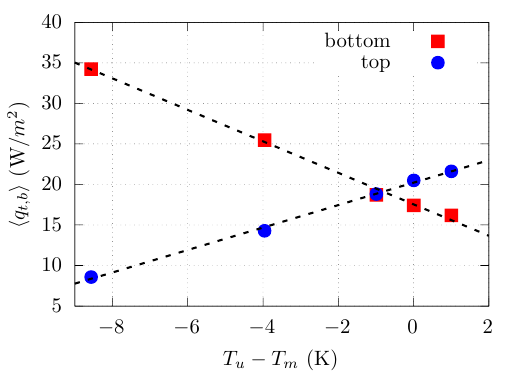}
    \caption{Effect of ambient temperature towards global heat flux measurement showing heat fluxes measured at the bottom (red squares) and top (blue bullets) plates and averaged over the entire horizontal length as a function of the difference between the temperature inside the Uboot ($T_{u}$) and mean temperature inside the cell ($T_m$).
	}
    \label{fig:varyingAmbientTemperature}
\end{figure}

Precise heat transfer measurements mandate good thermal insulation of the sidewalls of the RBC cell.
However, owing to the large dimensions of the experimental cell and the requirement that only
open-cell foam insulation material can be used under high-pressure conditions, flow through the
porous insulation material is unavoidable. This flow influences the temperature and velocity field within the cell, particularly due to the small aspect ratio, $\Gamma_y$. The influence of this effect is quantified through heat flux measurements performed at different mean temperatures, $T_m$. Figure~\ref{fig:varyingAmbientTemperature} presents the spatially and temporally averaged heat fluxes at the bottom plate (red squares) and top plate (blue circles) assemblies as a function of the temperature difference between the Uboot and the mean cell temperature, i.e., $T_u-T_m$. A clear dependence of the heat flux at both plates on this temperature difference is observed. For instance, when the Uboot is colder than the mean cell temperature (left side of the figure), a fraction of supplied heat is transferred from the cell to the Uboot and consequently does not reach the top plate assembly. As a result, heat flux entering the cell through the bottom plate assembly is significantly larger than the heat flux transferred from the convecting fluid to the top plate assembly. Conversely, when the Uboot is warmer than the cell, heat flux at the bottom plate assembly is smaller than that at the top plate assembly, as additional heat is introduced into the cell through the sidewalls. An approximately linear relationship is observed, suggesting that this heat loss is caused by creeping flow through the porous insulation material. It is assumed that the influence of sidewalls is minimized when heat fluxes at the top and bottom plate assemblies are equal. Therefore, in all subsequent experiments, $T_m$ is adjusted prior to each measurement such that the spatially averaged heat fluxes at the top and bottom plate assemblies are nearly equal.

\subsection{Global heat flux scaling}

\begin{figure}
    \includegraphics[width=\columnwidth]{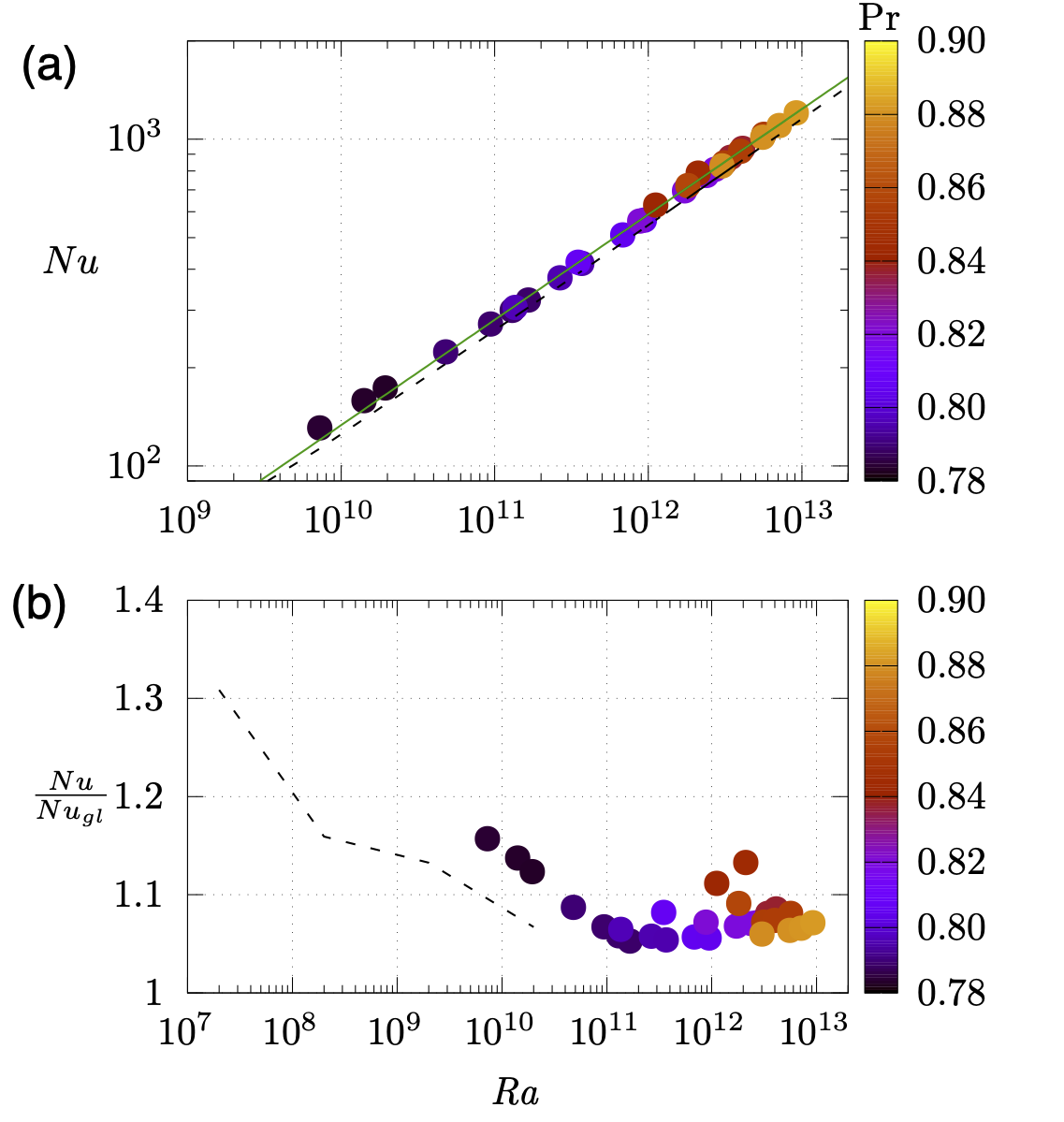}
	\caption{(a) \Nu\ as a function of \Ra\ on a log-log plot with \Pran\ represented by the
	colorbar. Note: green solid line marks a power-law fit to the data with $\Nu\propto\Ra^{0.32}$
	and dashed line marks predictions for \Nu\ based on the Grossmann-Lohse model for \Pran=0.82. \cite{Stevens.ea2013} (b) The same data normalized by the Grossmann-Lohse predictions ($\Nu_{gl}$). Note: dashed line marks results from DNS for a cylindrical system with isothermal sidewalls.\cite{stevens2014sidewall}
	}
    \label{fig:Nu-Ra}
\end{figure}

Fig.~\ref{fig:Nu-Ra} presents \Nu\ measurements as a function of \Ra\ as a benchmark test. These
measurements are obtained by conducting experiments using pressurized SF$_6$ at pressures ranging
from $1$ to $15\,$bar to realize Rayleigh numbers within the range $10^{10}\le \Ra\le 10^{12}$.
Owing to the variation in pressure, the Prandtl number also varies slightly, spanning the range
$0.78\le \Pran \le 0.88$. Figure~\ref{fig:Nu-Ra} displays the measured \Nu\ values, with the color
of the legend indicating the corresponding \Pran, which increases with increasing pressure. The Nusselt number is calculated from heat fluxes that are averaged horizontally over both the top and bottom plate assemblies and subsequently averaged in time. 

As shown in Fig.~\ref{fig:Nu-Ra}(a), the measured dependence $\Nu(\Ra)$ follows a nearly linear
trend on the logarithmic scale, indicating a power-law relationship of the form $\Nu\propto
\Ra^\gamma$. A fit to the data yields an exponent of $\gamma = 0.32$, represented by the solid green
line. Previous experimental and theoretical studies have demonstrated that $\gamma$ depends on both
\Pran\ and \Ra, and therefore varies over wider ranges of these control parameters. For comparison,
predictions of the Grossmann-Lohse (GL) model \cite{Stevens.ea2013} for $\Pran = 0.8$ are included
in Fig.~\ref{fig:Nu-Ra}(a) as a dashed black line. Although an overall good agreement is observed
between the measurements and the GL model, the measured values are consistently higher than their
$\Nu_{gl}$ counterpart by approximately $5-10\%$. This discrepancy is considered relatively small,
particularly given that the parameters of the GL model are determined from fits to other experimental data. Since most of these reference measurements are obtained through experiments performed in cylindrical cells with an aspect ratio of $\Gamma = 1$, different flow structures are expected and the influence of sidewalls is anticipated to be weaker than that in the present rectangular cell, which is characterized by a large horizontal extent ($\Gamma_x=5$) and a small transverse aspect ratio ($\Gamma_y=0.5$).

Fig.~\ref{fig:Nu-Ra}(b) shows the data normalized by the GL model prediction, i.e., $\Nu/\Nu_{gl}$.
As apparent from the figure, deviation from the GL model predictions becomes more pronounced with
decreasing \Ra\ (with measured heat flux at larger \Ra\ exceeding the GL model prediction by only about 5\%). It can therefore be hypothesized that the observed discrepancy may be due to sidewall effects since a similar increase in the discrepancy with a decrease in \Ra\ has also been reported in DNS with sidewalls maintained at constant temperature \cite{stevens2014sidewall} (shown by a dashed line in Fig.~\ref{fig:Nu-Ra}b). Typically, higher heat flux values are expected in the case of isothermal sidewalls, since heat can leave the cell near the hot bottom plate to the sidewalls and subsequently re-enter the system near the cold top plate. This additional heat transport is superimposed on the heat transported through convective flow within the cell. Stevens et al. \cite{stevens2014sidewall} demonstrated that this bypass mechanism occurs predominantly in the vicinity of the thermal boundary layers. Consequently, the effect is expected to become less significant with increasing $\Ra$, as the thermal boundary layers become thinner. Similar trends shown by the present experimental measurements and the DNS results obtained with isothermal sidewalls (dashed black line) support this interpretation. For the majority of the present measurements, however, this effect remains trivial. 

\section{Conclusions and Prospects}

A novel large-scale convection setup is designed to investigate Rayleigh-B\'enard convection at high Rayleigh numbers and large horizontal aspect ratios. The central component consists of a rectangular convection cell with a length of $3.5\,$m, a width of $0.35\,$m, and a height of $0.7\,$m, the latter of which can be readily adjusted. This cell enables accurate measurements of heat flux at both the top and bottom plate assemblies. It is installed inside the G\"ottingen Uboot facility and filled with SF$_6$ and pressurized up to $18\,$bar. This helps realize Rayleigh numbers of up to $\Ra=5\times 10^{12}$.

In contrast to previously employed geometries (such as cylindrical cells), the present design features a relatively large aspect ratio in one horizontal direction, i.e., $\Gamma_x = 5$, thereby providing sufficient lateral extent for the large-scale circulation to develop flow structures beyond a single convection roll. In addition, the experimental setup permits heat flux measurements with high spatial resolution along the $x$-axis. The validity of the apparatus is verified to be well suited for precise heat flux measurements and the results obtained are found to be in good agreement with previously reported findings.

In addition to its primary purpose of studying Rayleigh-B\'enard convection, the facility also enables investigations of horizontal convection since the top and bottom plates are divided into four laterally segmented sections that can be temperature-controlled independently. For this purpose, bottom-plate heating must be deactivated, and only the top-plate temperature is actively controlled. Temperature regulation is achieved via water circuits that operate independently for each of the four segments. While in classical RBC, the water is typically cooled to maintain the top plate at a lower temperature than the bottom plate, the connected chillers, in combination with heat exchangers, also allow the water to be heated without significant modification. There are two possible configurations using this facility to realize horizontal convection: (i) one is to heat one segment while cooling the one farthest from it and (ii) a superior alternative (similar to that investigated numerically by Reiter and Shishkina \cite{Reiter.Shishkina.2020}), wherein two outer segments are heated to generate stable stratification near the sidewalls and the two central segments are cooled, thereby introducing buoyancy forcing in the interior region.

The main focus of future research using this setup is directed towards the organization and dynamics of the large-scale flow pattern, which is to be inferred from temperature measurements at the center of the cell. In addition, the transparent sidewalls enable optical velocity measurements, such as particle image velocimetry or Lagrangian particle tracking, which are planned for future implementation.

\section{Acknowledgments}
The technical assistance from Artur Kubitzek, Andreas Renner, and Andreas Kopp in realizing the experimental setup presented in this manuscript is gratefully appreciated. SW acknowledges the support from the Max Planck University Twente Center for Complex Fluid Dynamics. UM acknowledges the support from the Max Planck Society through the Max Planck India Mobility Grant (2024-2026) which facilitated this collaboration.

\section*{Data Availability Statement}

The data that support the findings of this study are available from the corresponding author upon reasonable request.


\nocite{*}
\bibliographystyle{unsrt}
\bibliography{LHPCF.bib}

\end{document}